**Unlocking Electro-Optic Tuning in Hybrid Silicon Photonics Based on Ferroionic 2D Materials**

Ghada Dushaq [1]*, Solomon Serunjogi[1], Srinivasa R. Tamalampudi[1], and Mahmoud Rasras[1,2]*

[1] Department of Electrical and Computer Engineering, New York University Abu Dhabi, P.O. Box 129188, Abu Dhabi, United Arab Emirates.

[2] NYU Tandon School of Engineering, New York University, New York, USA

ghd1@nyu.edu, mr5098@nyu.edu

**Abstract**

Tunable optical materials are indispensable elements in modern optoelectronics, especially in integrated photonics circuits where precise control over the effective refractive index is essential for diverse applications. Two-dimensional materials like Transition Metal Dichalcogenides (TMDs) and graphene exhibit remarkable optical responses to external stimuli. However, achieving distinctive modulation across a broad spectrum while enabling precise phase control at low signal loss within a compact footprint remains an ongoing challenge. In this work, we unveil the robust electro-refractive response of multilayer ferroionic two-dimensional $CuCrP_2S_6$ (CCPS) in the near-infrared wavelength range. By integrating $CuCrP_2S_6$ into SiPh microring resonators (MRR), we enhance light-matter interaction and measurement sensitivity to minute phase and absorption variations. Results show that electrically driven Cu ions can tune the effective refractive index on the order of 2.8 x$10^{-3}$ RIU (refractive index unit) while preserving extinction ratios and resonance linewidth. Notably, these devices exhibit low optical losses of 2.7 dB/cm and excellent modulation efficiency of 0.25 V.cm with a consistent blue shift in the resonance wavelengths among all devices. These results outperform earlier findings on phase shifter based on TMDs. Consequently, CCPS integration endows characteristics akin to those of high-index active dielectric materials. Moreover, we showcase the electro-optic tuning sensitivity to light polarization, opening avenues for versatile light manipulation. The dual optoelectronic and ionotronic capabilities of the two-terminal CCPS devices hold vast potential, spanning applications such as phased arrays, optical switching, and neuromorphic systems in light-sensitive artificial synapses.

**Introduction**

The increasing demands for bandwidth in modern communication networks, optical sensing components, and advanced imaging systems have heightened the need for efficient tunable optical materials (TOM) designed for precise light modulation[1–4]. Two-dimensional (2D) materials have emerged as outstanding candidates for generating tunable optical components such as electro-optic modulators, switches, and filters, providing numerous possibilities for the precise and effective manipulation of light[5–10]. This attribute is a result of their quantum confinement, which allows for strong interactions between light and matter, leading to a pronounced change in their optical properties in response to external stimuli such as an electric field[11,12].

A variety of techniques have been developed to tune the optical properties of two-dimensional (2D) materials. For instance, successful demonstrations of light modulation in both the visible and short-wave infrared (SWIR) regions have been achieved through the utilization of electrostatic gating in transmission metal dichalcogenides (TMDs) when integrated into photonic circuits[7,13–17]. However, in the vicinity of their excitonic resonances, these materials exhibit simultaneous modulation of both the refractive index ($\Delta n$) and absorption ($\Delta k$). As a result, attaining pure phase modulation presents inherent challenges. Recently, a pure phase modulation based on integrated $WS_2$ and $MoS_2$ operating in the SWIR (far from their excitonic resonances) has been achieved[7]. The findings demonstrated a high $|\Delta n/\Delta k|$ ratio as well as minimal losses during signal transmission. Nevertheless, the realization of phase modulators predicated on capacitive configurations to facilitate electrostatic doping of $WS_2$ presents noteworthy challenges predominantly rooted in the intricacies of lithographic processes and precise alignment requirements. Additional approaches, such as the employment of the plasma dispersion effect in silicon-based devices, induce intensity change along with the phase shift[18–20]. Moreover, the use of thermo-optic methods poses significant challenges due to the need for efficient heat dissipation and the maintenance of temperature stability[21]. Hence, identifying new materials with unique modulation properties to control Δn/Δk across a broad bandwidth within a compact footprint is highly desired.

Very recently, there has been a significant surge in interest in 2D ferroionic compounds[22–24], primarily due to their remarkable characteristics such as moderate bandgaps ranging from 1.3 to 3.5 eV, multiple ferroic orders, and exceptional ionic conductivity[23,25–29]. Notably, Van der Waals (vdW) thio- and seleno-phosphates such as $CuInP_2S_6$ (CIPS) and $CuCrP_2S_6$ (CCPS) have found application as active dielectrics in electronic devices with two-dimensional or quasi-two-dimensional designs. The overall electronic attributes of these materials are heavily influenced by their bulk ionic conductivity, a critical aspect that can be effectively controlled by manipulating factors such as poling time, polarization, and current direction, all of which are connected to the migration of highly mobile copper (Cu) ions[25,30–33]. Despite the advantageous presence of electrochemically active Cu ions within the lattice and their facile integration into various substrates, current research predominantly focuses on their memristive properties for storage applications and neuromorphic computing[34–36]. However, by combining its multi ferric order and ionic attributes, $CuCrP_2S_6$ offers a novel pathway to address challenges in achieving pure optical phase modulation in integrated photonics while enabling efficient and versatile light manipulation.

In this study, the electro-optic response of multilayer CCPS integrated on SiPh microring resonators (MRR) are actively manipulated at the near-infrared (NIR) wavelengths. Electrically driven Cu ion migration enables precise tuning of the refractive index by approximately 2.8 x$10^{-3}$ RIU (refractive index units), while preserving extinction ratios and resonance linewidth. This adjustable electrical conduction arises from the reversible Cu ions accumulation or removal at the metal-semiconductor interface, modulating the contact Schottky barrier height. The inclusion of CCPS on uncladded MRR results in strong light-matter interaction and low optical losses of 2.7 dB/cm. This value is lower than barium titanate devices[37] and comparable to those of lithium niobate devices[38,39]. Additionally, the devices exhibit excellent modulation efficiency of 0.25 V.cm with a consistent blue shift in the resonance wavelengths among all devices. Electro-optic tuning exhibits sensitivity to the light polarization alignment with the CCPS, highlighting its significance. The combined optoelectronic and ionotronic functionalities in the two-terminal CCPS device hold potential for applications, including phased arrays, optical switching, and light-sensitive artificial synapses in neuromorphic systems.

**Results**

**Structural characteristics of the vdW CCPS crystal**

The layered Van der Waals $CuCrP_2S_6$ (CCPS) belongs to the transition metal thio/selenophosphates (TPS) family with a monoclinic crystal structure (Pc space group)[26]. Figure 1 shows the top view (ab plane) and side view (bc plane) of the material structure. The monolayer is composed of a sulfur framework in which octahedral cages are filled by Cu and chromium (Cr) ions as well as P-P pairs. The Cu ions occupy the upper (Cu1) and lower (Cu2) positions alternately, resulting in an antiferroelectric (AFE) state at low temperatures. Within the crystal structure, triangular networks are formed by quasi-trigonal $CuS_3$, octahedral $CrS_6$, and $P_2S_6$ units[26,28,40,41]. The Cr ions and the P-P pairs are almost centered within a layer, while the Cu ions are off-centered[42]. Interestingly, the Cu ions become mobile under an external electric field, and they exhibit in-plane and out-of-plane (across the Van der Waals gap) movements, which is the primary cause of the ionic conductivity in this layered material.

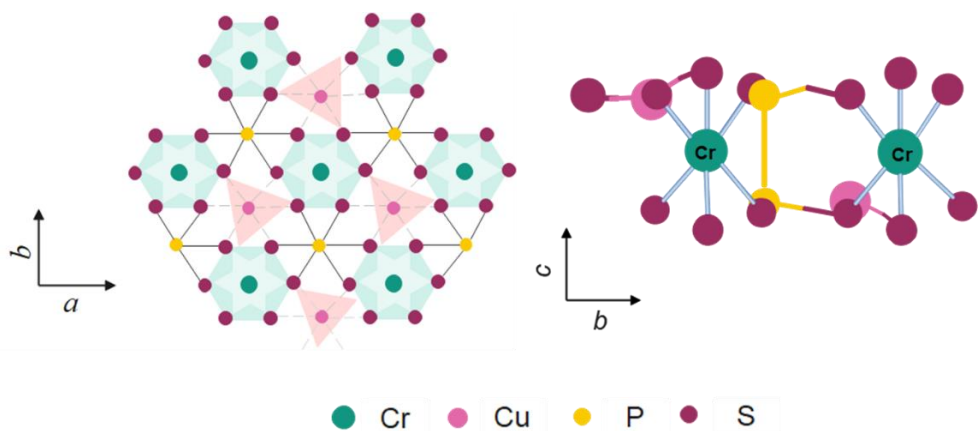

*Fig 1. **Schematic representation** of the top view (ab plane) and side view (bc plane) of CCPS*

To quantify the crystal structure, the CCPS flakes are directly exfoliated from their bulk crystals using mechanical exfoliation technique. To ascertain a good crystalline structure and homogeneity of the layered CCPS, transmission electron microscopy (TEM) testing was carried out. A low-resolution TEM image of an exfoliated flake is shown in Fig. 2a (a high-resolution image is shown in Fig.S1). Additionally, the selected area diffraction pattern (SADP) along the c-axis of the CCPS verified the monoclinic nature and the high consistency of the crystals (see Fig. 2a)[26]. Raman spectra collected from CCPS flakes in the backscattered geometry are shown in Fig. 2b. The four active vibrational modes at room temperature are labeled as I-IV. A comparison between our measured Raman peaks and those reported in the literature reveals that peak I corresponds to in-plane vibrational mode ($E_g$), while all other peaks correspond to out-of-plane mode ($A_g$)[43]. Quantitative elemental mapping using energy dispersive x-ray spectroscopy (EDX) is shown in Fig. 2d. The maps confirm the uniform distribution of the elements within CCPS; additionally, an atomic ratio of 16.8: 2.2: 4.7:12.9 is detected, which is almost stoichiometric with respect to CCPS (1:1:2:6), the high-count rate of copper is due to the TEM copper grid utilized for imaging in the transmission electron microscopy (the full EDX spectrum is shown in Fig.S2).

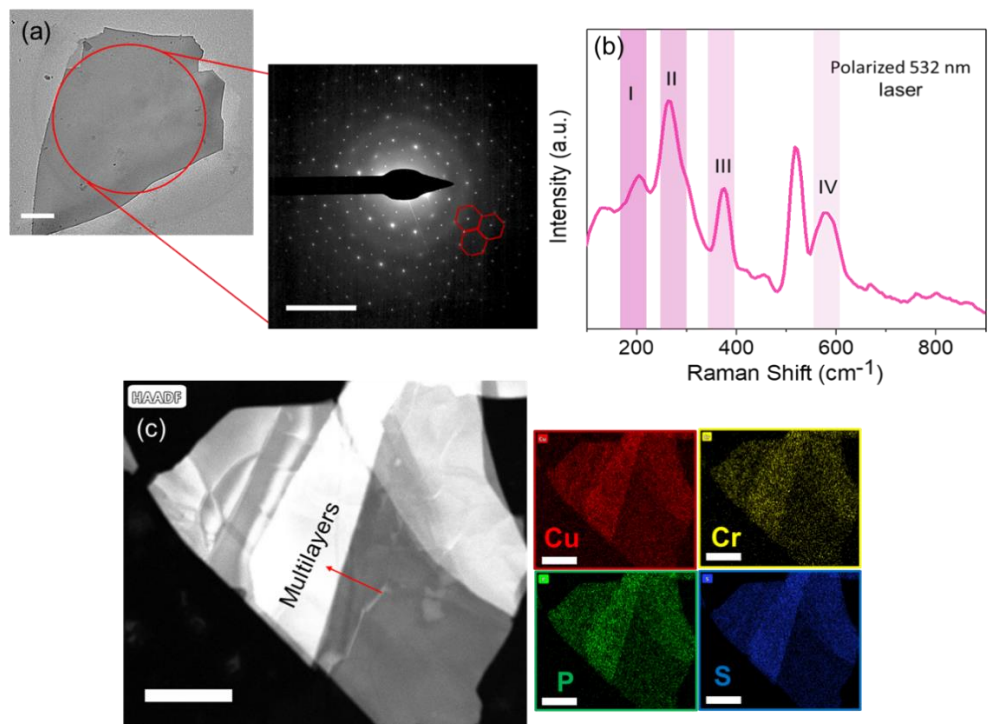

*Fig 2. **Structural characteristics of CCPS** (a) TEM image along with SADP, the red dashed hexagon is highlighted to guide the eye into the structure (b) Raman spectra collected using a polarized 532 nm laser (c) quantitative EDX mapping of CCPS. The bar readings for (a) and (c) are: 500 nm, 10 nm$^{-1}$ and 1 μm, respectively.*

**Electrical Modulation and Ionic Migration in CCPS**

Through systematic investigations of the ferroic compounds, $CuInP_2S_6$ and $CuCrP_2S_6$, it has been ascertained that the dominant determinant governing their electrical behavior originates from their profoundly mobile electrochemically active Cu ions within the lattice structure[25,30–33]. At ambient temperatures, these materials exhibit insulating characteristics under low applied voltage conditions. On the other hand, they demonstrate exceptionally sensitive current response to electric stimuli beyond a certain threshold voltage. This remarkable phenomenon of reversible switching between insulating and conducting states, combined with the graded local distribution of charged Cu ions, has been examined in this section. This electrical characterization bears significant advantages in facilitating the correlation between optoelectronic and ionotronic conduction processes, as further elucidated in the subsequent section dedicated to electro-optic tuning.

Figure 3a depicts the current-voltage characteristic (*I-V*) of a symmetric planar Au/CCPS/Au device, obtained with a ramping rate of 0.5 V/s. During the initial stage, a positive bias voltage is swept from 0 to 7 V, triggering a transition from a high resistive state (HRS) to a low resistive state (LRS) at a threshold voltage ($V_{th}$) of approximately 6.0 V. However, upon reversing the positive voltage sweep (from 7 V to 0V), the device reverts to the HRS. The presence of a hysteresis loop indicates a bidirectional threshold resistive switching (RS) effect, making the device a promising candidate for neuromorphic computing[44,45].

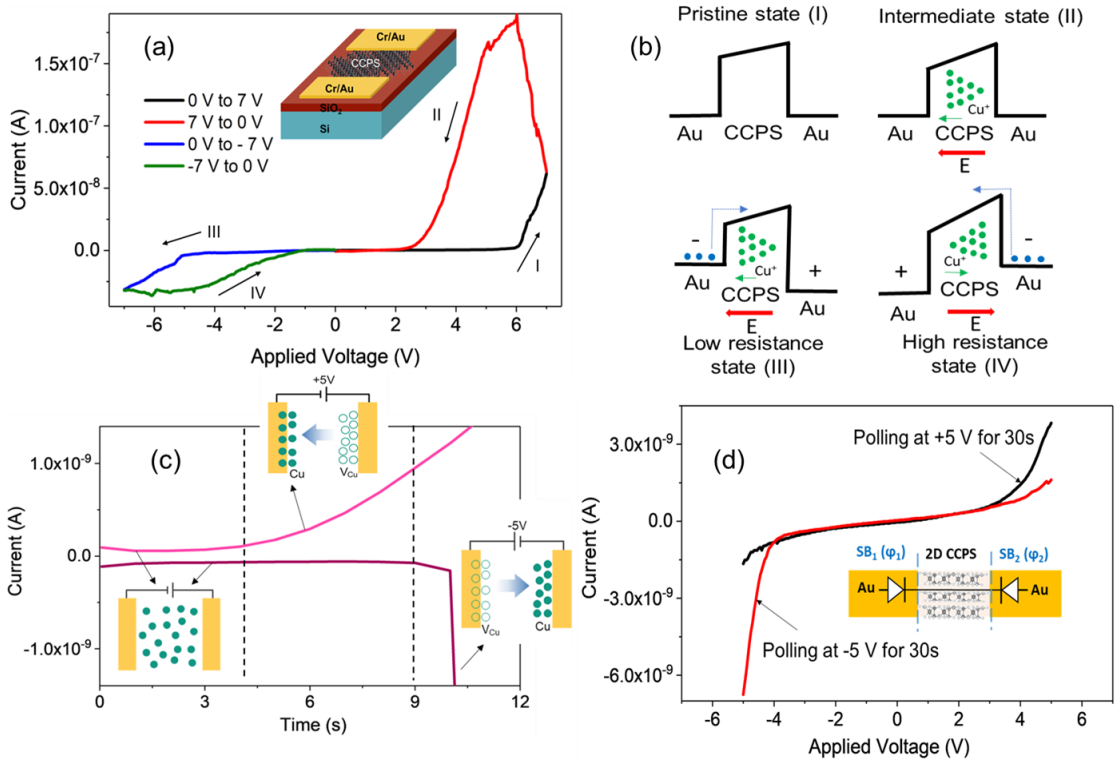

*Fig 3.* **Electrical characteristics of Au/CCPS/Au structure** *(a) full I −V curves swept in the order 0 V → 7 V → 0 V → -7 V → 0 V labeled as I, II, III, IV, inset shows a schematic of the fabricated planner Au/CCPS/Au structure (b) schematic of band alignment at different bias condition (c) current- time (I-t) characteristics at ± 5 V along with a schematic representation of Cu distribution (d) forward and*

*backward rectification states obtained by polling the voltage at ±5 V for 30 seconds and subsequently performed a complete cycle of the I-V characteristics.*

Additionally, when the *I-V* characteristics are cycled under negative bias voltage (0 to -7 V, and then from -7 back to 0 V), similar behavior is observed compared to the forward scans under positive bias voltages. To further investigate the RS characteristics, the *I-V* response is examined with different sweep voltages ranging from ±5 to ±10 V applied to the planar Au/CCPS/Au device, as shown in Fig S3. This result demonstrates a switchable bidirectional diode behavior accompanied by resistive switching for these voltage ranges. The switching ratio ($I_{LRS}/I_{HRS}$) reaches 0.1 µA/25 pA, equivalent to $4 \times 10^3$.

The observed behavior is in accordance with prior results obtained using a conductive scanning tip, where pronounced hysteresis and rectification behavior, contingent on sweeping speed, are attributed to the migration and uneven accumulation of Cu ions[25,29]. Importantly, this process exhibits reversibility, as long as the device remains undamaged by the substantial current induced by Joule heating.

Figure 3b depicts the band diagram of the Au/CCPS/Au device under various bias voltage conditions, labeled from I to IV. In the absence of applied bias, the Cu ions are uniformly distributed throughout the CCPS phase, representing the pristine state I. However, upon the application of bias voltage, the Cu ions undergo migration, leading to a nonuniform distribution compared to the unbiased condition, forming the intermediate state II.

States III and IV reveal the resistive switching and bidirectional rectification phenomena. In state III, the electric field progressively increases toward the left, accelerating the migration of Cu ions in the same direction, resulting in a higher Cu accumulation at the left Au electrode. As a consequence, this electrode attains n-type characteristics, while the other electrode experiences Cu ion depletion, leading to a p-type state. Consequently, the barrier height reduces due to the increased accumulation of Cu ions at the Au/CCPS interface, which is expected to enhance carrier transport across the interface. Similarly, a corresponding trend is observed during opposite polarity sweeping, as depicted in state IV.

To further investigate the forward and backward rectification states, we polled the voltage at ±5 V for 30 seconds and subsequently performed a complete cycle of the *I-V* characteristics. As demonstrated in Fig. 3d, the homojunction Au/CCPS/Au exhibits dual rectification, confirming the ionic conduction and redistribution of Cu ions during opposite polarity conditions.

Moreover, Fig. 3c and Fig.S3 illustrate the time-dependent conductivity of the Au/CCPS/Au structure. It is worth noting that the initial current of $2 \times 10^{-12}$ A exhibits a gradual increase within a few seconds to a few microamps. This observed current progression implies that the ionic conductivity of copper can be stimulated when an external electric field of sufficient strength overcomes the potential barrier. The depletion of Cu ions from the lattice results in the creation of additional vacancies, thereby elevating the possibility of ion hopping and diffusion. Consequently, the ionic conductivity experiences exponential growth over time. This trend remains consistent when the voltage polarity is reversed. Additionally, the voltage ramping rate in V/s affects the amount of the loops observed in the *I-V* characteristics as depicted in Fig.S3.

**Photonic Design and CCPS Integration: Theory and Simulation**

Figure 4 depicts a schematic configuration of the hybrid integration of multilayer CCPS into the silicon photonics (SiPh) circuits. In this study, SOI wafers with a 220 nm top silicon layer and a 2 µm buried oxide layer are used. The MRR cavity's waveguide width (d) is 460 nm, and its ring radius is 45 µm. A 100 nm gap spacing is used to couple light to the micro-resonator from a bus waveguide and collected at the through port output.

Incorporating a CCPS material on top of the ring waveguide offers the ability to control light propagation in the "add" and "through" ports of the resonator by adjusting the real part of the CCPS's refractive index via electro-optic tuning. Light passing through the waveguide is coupled to the top CCPS layer evanescently, where the coupling strength is influenced by the thickness of the CCPS. The expressions for the transmission of the resonator's exit ports can be formulated as follows[46]:

$$T_{through} = \frac{t^2\alpha^2 - 2t_1 t_2 \alpha \cos\theta + t_1^2}{1 - 2t_1 t_2 \alpha \cos\theta + (t_1 t_2 \alpha)^2} \quad (1)$$

$$T_{drop} = \frac{(1-t_1^2)(1-t_2^2)\alpha}{1 - 2t_1 t_2 \alpha \cos\theta + (t_1 t_2 \alpha)^2} \quad (2)$$

Where α is the attenuation factor, θ is the phase factor, t₁ and t₂ are coupling parameters, α, and θ can be expressed as:

$$\alpha = \exp(-\tfrac{2\pi}{\lambda}[k_{eff,wg}(2\pi R - L_{CCPS}) + k_{eff,CCPS} L_{CCPS}]) \approx \exp(-\tfrac{2\pi}{\lambda} k_{eff,CCPS} L_{CCPS}) \quad (3)$$

$$\theta = \tfrac{2\pi}{\lambda}[n_{eff,wg}(2\pi R - L_{CCPS}) + n_{eff,CCPS} L_{CCPS}] \quad (4)$$

The effective imaginary and real components of the refractive index of the Si waveguide with (without) CCPS are denoted by $k_{eff,CCPS}$ ($k_{eff,wg}$) and $n_{eff,CCPS}$ ($n_{eff,wg}$), respectively. R is the radius of the ring waveguide, while $L_{CCPS}$ is the length of the integrated CCPS flake.

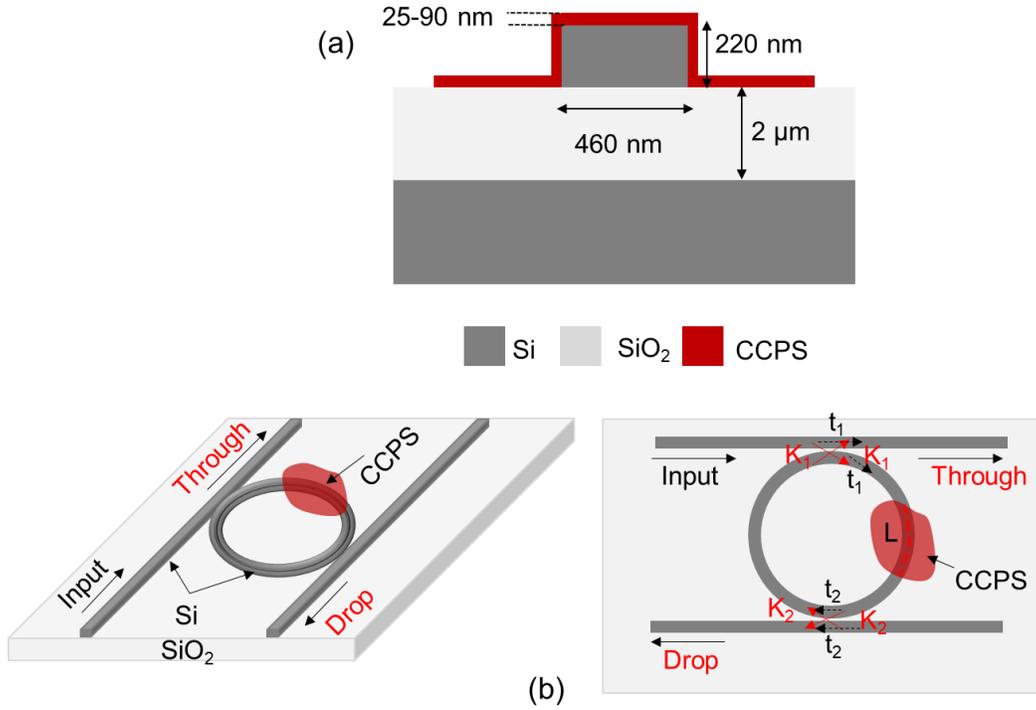

*Fig 4. **Schematic configuration of the photonic design and CCPS integration** (a) cross section showing design parameters (b) 3D and cross-section of the Si MRR design with integrated CCPS.*

The optical parameters, namely, refractive index ($n$) and extinction coefficient ($k$) of exfoliated CCPS on 300 nm SiO$_2$/Si substrate are extracted from high-resolution spectroscopic ellipsometry measurements (more details on ellipsometer measurements, and n, k values are included in the supplementary material). The extinction coefficient is linked to the absorption coefficient ($\alpha$) as $\alpha=4\pi k/\lambda$, hence, the absorption coefficient can be calculated and fitted to Tauc's equations, as shown in Fig.5a. The extracted indirect bandgap is about 1.35 eV, whereas the direct bandgap is higher by 0.9 eV. It is worth mentioning that no photoluminescent emission was observed at room temperature. Furthermore, these values are consistent with the literature and the low optical loss (transparency) in the NIR range (1280 nm -1360 nm) as will be discussed in the next section[25].

The interaction between guided light and the multilayer-CCPS plays a pivotal role in designing optimal optoelectronic devices on the SiPh platform[47]. Consequently, numerical and experimental investigations were conducted to test the variations in the effective refractive index ($n_{eff}$) and optical losses of the hybrid Si/CCPS structures. These calculations were executed for different CCPS thicknesses, with a specific focus on the NIR wavelength from 1280 nm to 1360 nm. The range of CCPS thickness (30 nm to 90 nm) used in the simulation is aligned with our experimental tests. The measured values of refractive index ($n$) and extinction coefficient ($k$) of multilayer CCPS were then fed into the mode analysis of Lumerical Mode Solver Software.

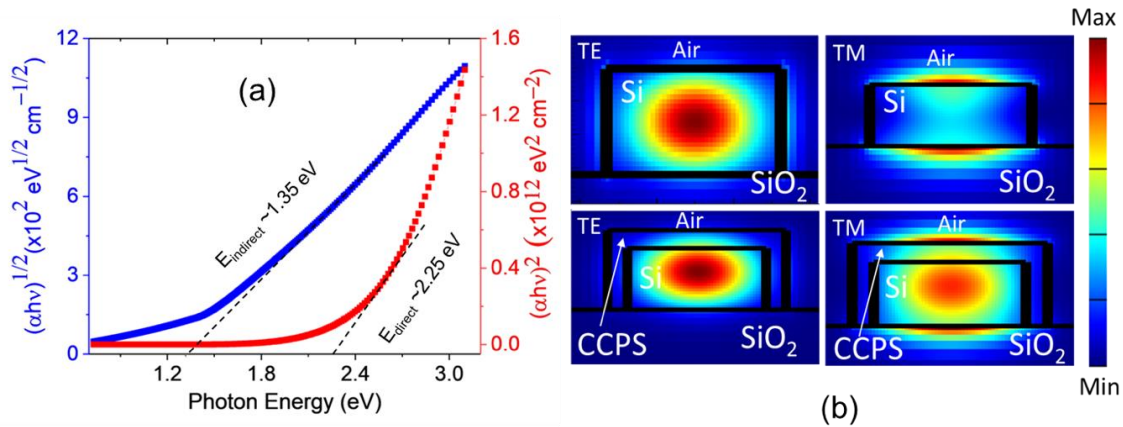

*Fig 5. **Optical properties and guided light/ CCPS interaction** (a) direct and indirect transition Tauc plots of the absorption coefficient derived from high-resolution spectroscopic ellipsometry (b) Electric-field profiles ($|E|^2$) of TE and TM modes of bare Si waveguide (top panel) and 65 nm CCPS on Si (bottom panel) at 1310 nm.*

In Figure 5b, the electric-field profiles ($|E|^2$) of the fundamental guided modes are displayed for both a bare silicon waveguide and integrated structures with a 65 nm CCPS layer at 1310 nm. As observable, the silicon waveguide supports quasi-TE (transverse electric) and quasi-TM (transverse magnetic) modes.

Furthermore, the mode evanescent field effectively overlaps with the multilayer CCPS flake for both polarizations, where the fraction of the optical mode power in CCPS is ~11% and ~12% for TE and TM polarization, respectively. The mode profile and overlap factor for other CCPS thicknesses are included in Fig.S5

**Device Fabrication and Testing**

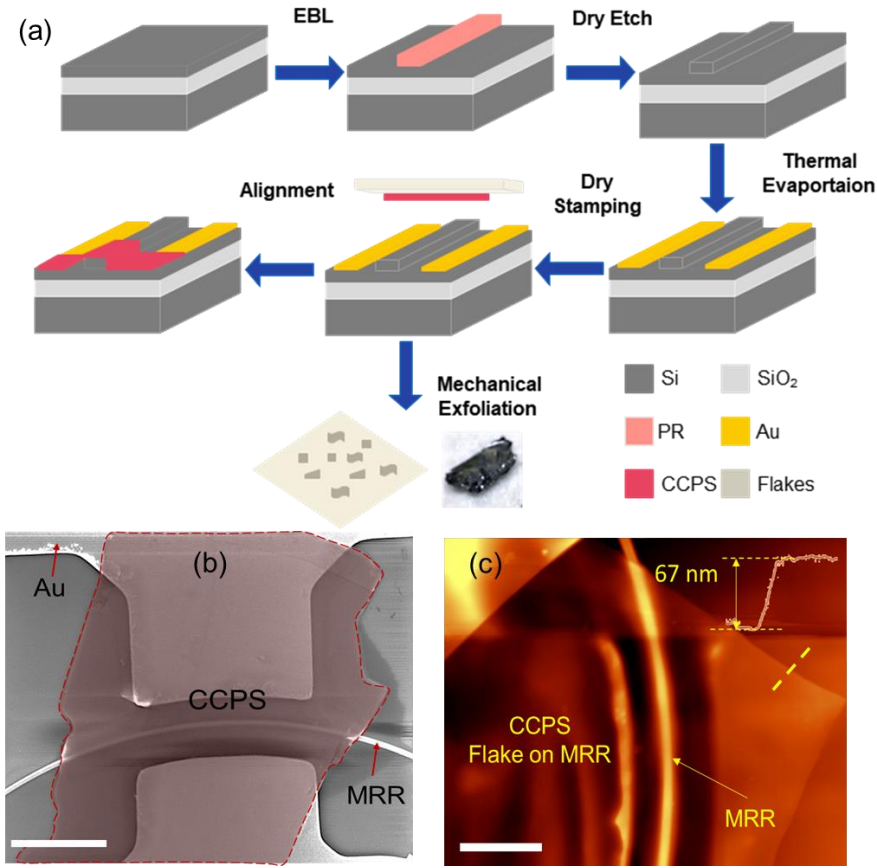

*Fig 6. **CCPS integration on SiPh platform** (a) schematic representation of the fabrication and CCPS integration process flow (b) SEM image of the transferred CCPS on MRR (c) AFM image scan, the dashed yellow line shows a thickness of ∼ 67 nm CCPS. Bars in (b) and (c) are 10 μm and 6 μm, respectively.*

The multilayer CCPS flakes were obtained through dry mechanical exfoliation from their bulk crystals, which were commercially available from HQ Graphene. In this investigation, we systematically transferred CCPS flakes with varying thicknesses ranging from 32 nm to 100 nm onto the MRR structure utilizing a deterministic dry transfer process[48–50]. Fig.6a shows a schematic representation of the fabrication and CCPS integration process. The hybrid Si/CCPS MRR's morphology is depicted in Fig. 6b through scanning electron microscope (SEM) images. Evidently, the CCPS flakes exhibit good alignment and conform to the photonic structure beneath, facilitating strong light/CCPS coupling. To quantitatively assess the thickness of the transferred flakes, atomic force microscopy (AFM) was employed, and the results are depicted in Fig. 6c. Additionally, Fig S6. displays a 3D reconstructed AFM scan, showcasing the conformal coverage and strong adhesion of the CCPS flakes on both the top surface and sidewalls of the MRR waveguide.

**Optical Transmission Measurements**

To experimentally verify the modulation of the hybrid integrated waveguide's refractive index and the optical losses, the optical transmission spectra of the CCPS-integrated MRR were measured. For this study, we used a tunable laser operating at SWIR wavelengths to edge-couple transverse electric (TE) polarized light into the MRR through a lensed fiber. The output response was collected by another lensed fiber and detected using a power meter. The light in the Si waveguide is evanescently coupled into the CCPS flake. Figure 7a displays the resonance dips of the MRR before (described as reference) and after integrating a CCPS flake with a thickness of ~ 65–67 nm and a coverage length of about 30 μm. The corresponding response for the TM mode can be found in the supplementary information Fig. S7. Additionally, we examined the MRR's response at various CCPS coverages and included their transmission spectra in the supplementary information Fig. S7. Intriguingly, the incorporation of CCPS into the MRR structure results in strong light-matter interaction with ~ 5 dB higher dips extinction ratio (ER) compared to bare silicon and low optical losses (see Fig.7a and 7b).

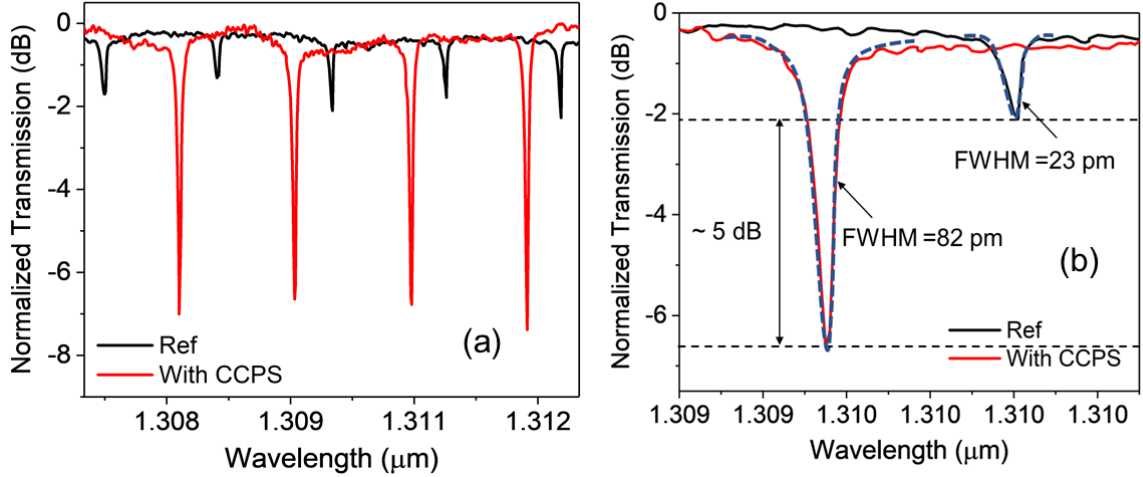

*Fig 7. **Passive optical measurements** (a) Measured transmission spectra of TE mode in Si/CCPS hybrid MRR with a radius of 45 μm and CCPS thickness of ~ 67 nm a long with bare Si-MRR as a reference (b) zoom-in image of the MRR resonances with and without CCPS showing the change in FWHM of the resonance dips, dashed line is the Lorentz fitting while solid line is the experimental transmission spectra.*

Moreover, we experimentally extracted the optical losses due to the multi-layer CCPS by analyzing the variation in the MRR's Q-factor[51]. As illustrated in Fig. 7b, the loaded quality factor ($\lambda_0$/FWHM) decreased from ~ $5 \times 10^4$ (for the bare Si MRR, the value averaged over 5 tested rings) to ~ $2 \times 10^4$ (for the hybrid CCPS/Si MRR with a flake thickness of 67 nm and interaction length of 30 μm, the value averaged over 5 CCPS loaded rings). This observation highlights the low-loss characteristics of CCPS in the SWIR wavelength. To evaluate the additional optical losses induced by the integrated CCPS flakes in the TE polarization, we calculated the difference between the loss of the reference ring and that obtained in the hybrid structure (CCPS/Si). These losses primarily originate from reflections and scattering at the coupling interface between the passive waveguide (air/Si) and the hybrid region, involving mode mismatching losses and flake irregularities. Negligible losses are attributed to material absorption in the SWIR.

The extracted optical losses amount to 2.7 dB/cm. Further details on the calculation are available in the supplementary information. This value is lower than barium titanate devices[37] and comparable to those of lithium niobate devices[38,39] and monolayer TMDs optimized to operate on their transparency window (O and C optical bands) for different near-infrared electro-optic applications[7] ( see table 1).

**Electro-Optic Tuning**

To probe the ionic migration-induced electro-optical response of CCPS, a hybrid Si–CCPS MRR platform was utilized. A 20-35 μm arc length of CCPS was integrated onto the ring. As depicted in Fig.6b, the flakes are stamped into pre-patterned 100 nm Au/10 nm Cr electrodes separated by 3.25 μm on both sides of the waveguide. This cavity is critically coupled into a Si bus waveguide, making the ring's transmission spectrum highly sensitive to minute phase and absorption variation within the cavity. We tune the effective index of the optical mode by harnessing the migration of Cu ions driven by the electric field within the CCPS, as illustrated in Fig S8 of our experimental setup. Fig. 8a illustrates the impact of an applied potential difference across the electrodes connected to the CCPS, resulting in a progressive redistribution of Cu ions. This phenomenon consequently instigated a discernible alteration in both the carrier concentration and the material's electrical conductivity. Due to the soft nature of Cu–S bond, it allows the Cu ion hop between the intralayer and interlayer sites and even across the vdW gap interlayer under the influence of an electric field (see Fig.1). Hence, optical tuning can be achieved by utilizing the reversible process of Cu ions extraction from sulfur octahedra through van der Waals gaps and re-intercalation into layers, a process reliably controlled by the applied voltage.

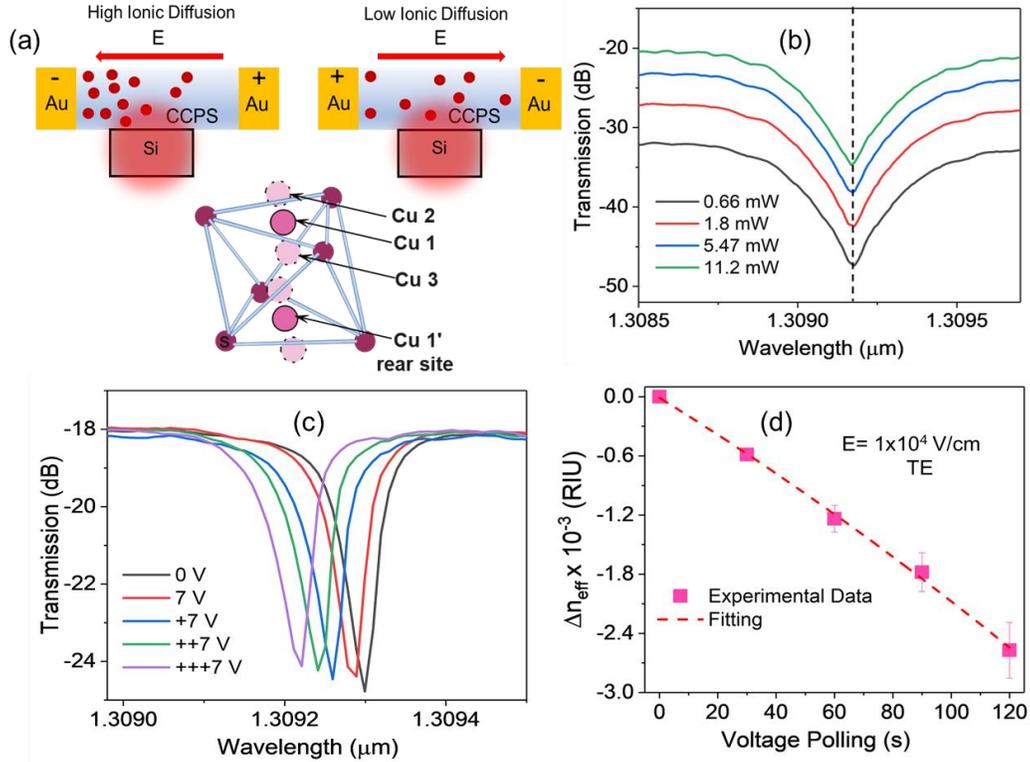

*Fig 8. **Electro-optical testing of hybrid CCPS/Si MRR** (a) Visualized illustration of copper redistribution in CCPS/Si heterostructure upon application of electric field along with the movements of Cu in the sulfur cage (b) the transmission spectra of MRR at different input optical power (c) the transmission spectra of MRR at a constant voltage of 7V as the polling time progressed (d) the change in effective index of refraction versus the polling time at E= $1\times10^4$ V/cm.*

The effect of the applied bias on resonance wavelength shift was carried out at SWIR. To mitigate any potential influence from the laser input power, the optical power delivered to the Si/CCPS chip was systematically varied from 0.66 mW (-1.8 dB) to 11.2 mW (10.5 dB), while maintaining the device bias at 0V. As evident in Fig. 8b, there is no discernible change in the positions of the resonance peaks for both polarizations. Accordingly, the propagating light induces no noticeable thermal dissipation in the device. Thus, we limited the laser power in all subsequent measurements to less than 11.2 mW (10.5dB). Note that the 10.5 dB is the laser input power which is not the power delivered to the CCPS/Si-MRR. The power delivered to CCPS/Si-MRR is -1.5 dB [10.5 dB (laser input power)-12 dB (insertion loss per facet)]

Figure 8c depicts the transmission spectra for TE polarization at bias voltages of 0 V and 7 V. The spectra were recorded with a constant voltage of 7V as the polling time progressed. The polling instances are labeled as +7, ++7, +++7, spaced 30 seconds apart from each scan. Across all scans, a consistent blue shift in the resonance wavelengths was observed for all fabricated devices (see Fig.S9). This blue shift implies an electro-optical modulation of the refractive index of the combined Si/CCPS guiding structure, given that thermal heating typically results in a red shift in the Si-MMR resonance wavelength[52]. Remarkably, the application of a bias across the device does not affect the extinction ratios and the resonance linewidth. Therefore, the active migration of Cu ions has no impact on the imaginary part of the refractive index. In contrast, plasma dispersion effect-operating devices frequently exhibit amplitude chirp during active operation due to undesired optical absorption[19].

Fig. 8d depicts the change in the effective index of refraction as a function of polling time at 7V (E=V/d). Figure. S10 shows the resonance shift as a function of increasing voltage, from this, we estimate a tuning efficiency ($\delta\lambda/\delta V$) of -8.3 pm/V. The active migration of Cu ions demonstrates the ability to electrically tune the effective refractive index on the order of $2.8 \times 10^{-3}$ RIU (see supplement note 2). Moreover, the half-wave voltage-length product is $V_\pi L = 0.5 \cdot L \cdot FSR/(\delta\lambda/\delta V) \approx 0.25$ V.cm ( FSR=1.4 nm, L=30 µm) which is lower than the phase shifter demonstrated based on TMDs[7,16,17] ( see table 1).

Figures 9a and 9b show the transmission spectra for the TE and TM modes, respectively, under varying polling times at an applied voltage of 6V. This voltage induces the migration of Cu ions in the integrated CCPS/Si MRR. In the case of the TE mode (Fig. 9a), a pronounced wavelength shift is observable, while the TM resonance mode (Fig. 9b) displays no discernible shift. This observation suggests that the influence of ionic conductivity and resistance change is significant only when the optical electric field aligns with the CCPS. Specifically, this effect is pronounced when the $E_x$ component in TE polarization aligns with the applied electric field, as illustrated in the schematic representation in Figures 9a and 9b.

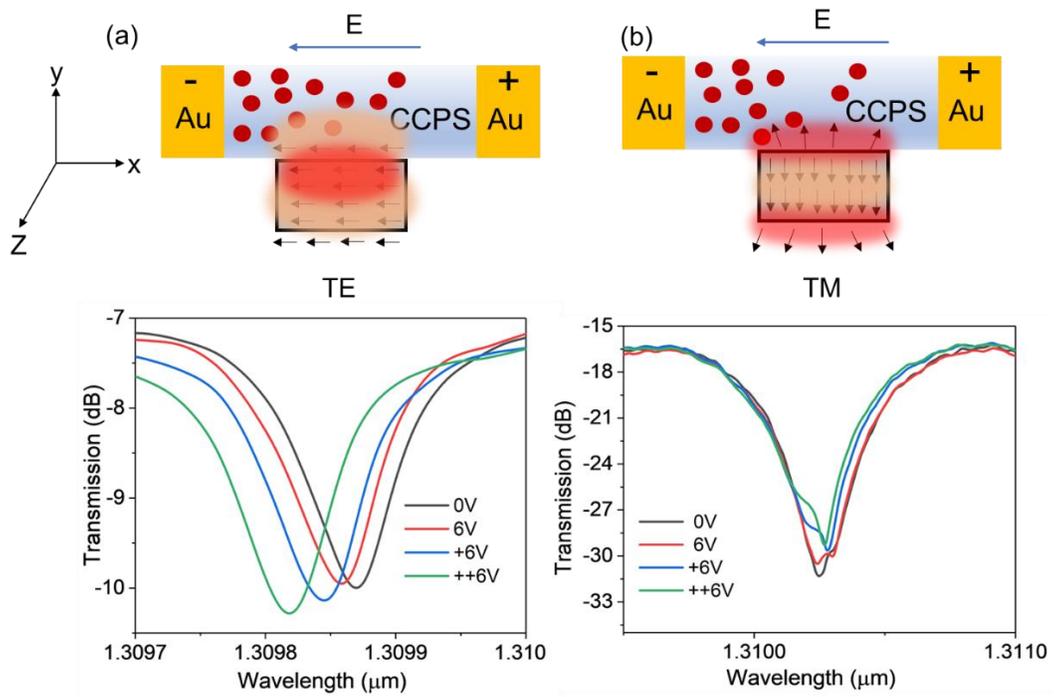

*Fig 9. **Electro-optic response of the MRR** for the (a) TE mode (b) TM mode at a constant voltage of 6V as the polling time progressed along with graphic representation showing the direction of electric field components for both polarizations.*

**Table 1** Summary of electro-optic phase shifter fabricated based on ferroelectric materials (BaTiO$_3$, LiNbO$_3$) and electrostatic doping of TMDs.

| Material platform | Photonic structure | Length of modulation area (μm) | Tuning efficiency (pm/V) | $V_\pi \cdot L$ (V.cm) | Optical propagation loss (dB/cm) | Ref |
|---|---|---|---|---|---|---|
| **LiNbO$_3$ (LN)** | Si/LN Mach–Zehnder modulators (MZMs) | 5000 | - | 2.2 | 0.98 | *Nat. Photonics* **13**, *359–364, 2019* |
| **BaTiO$_3$ (BTO)** | Hybrid BTO-SiO$_2$ racetrack resonator on SiPh | 100 | 923 | 0.45 | 10.5 | *Nat. Mater.* **18**, *42–47, 2019* |
| **BaTiO$_3$ (BTO)** | BTO-SiN racetrack resonator | 150 | - | - | 9.4 | *ACS Photonics* **6**, *2677–2684, 2019* |
| **TMD** | Monolayer WS$_2$ on SiN-MZI | 500 | - | 0.8 | 1.5 | *Nat. Photonics* **14**, *256–262, 2020* |
| **TMD** | Few-layer MoS2 on SiN micro-ring resonator- | - | 29.42 | 0.69 | - | *Photonics Res.* **10**, *1401–1407, 2022* |
| **Ferroionic 2D materials** | Multilayer CCPS on mirroring resonator | 30 | 8.3 | 0.25 | 2.7 | **_This work_** |

**Conclusion**

In summary, we demonstrated a novel avenue for active light manipulation through the utilization of ferroionic 2D CCPS material. When integrated onto SiPh microring resonators, these materials exhibit a remarkable ability to finely tune the effective index of refraction without any amplitude chirp. This is attributed to the adjustable electrical conduction that originates from the reversible accumulation or removal of mobile Cu ions at the metal-semiconductor interface. Results show that electrically driven Cu ions can tune the effective refractive index on the order of 2.8 x$10^{-3}$ RIU (refractive index unit) while preserving extinction ratios and resonance linewidth. Additionally, the devices feature remarkably low optical losses of 2.7 dB/cm and excellent half-wave voltage-length product of $V_\pi L = 0.25$ V.cm which is lower than the phase shifter demonstrated based on TMDs. We further determine the electro-optic tuning sensitivity to light polarization. It is evident that the influence of ionic conductivity and resistance change is significant only when the optical electric field aligns with the Cu ions migration direction (same as the applied field). Strong light-matter interaction and active dielectric properties of CCPS offer a new pathway to address challenges in achieving pure phase modulation while enabling efficient and versatile light manipulation.

**Experimental Setup**

**Materials supply**

$CuCrP_2S_6$ bulk crystals were purchased from a hq graphene vendor (https://www.hqgraphene.com/).

**Scanning electron microscopy (SEM)**

Photonic chips were mounted on a SEM stub using carbon tape and imaged under high vacuum mode by using a (FEI) Quanta 450 field emission scanning electron microscope with an electron energy of 10 KV.

**Atomic force microscopy (AFM)**

It was performed using a WITec Atomic Force Microscope (AFM) module integrated with a research-grade optical microscope in the tapping mode. The cantilever tip (Scanasyst-air) had a radius of 7 nm, a force constant of 0.2 N/m, and a resonance frequency of 14 kHz.

**Raman spectroscopy**

Confocal micro-Raman (WItec alpha 300) with a laser excitation source of 532 nm was precisely subjected to the flake using a 100× objective lens with a spot size of ∼0.7 μm. To avoid any damage to the sample, a low laser power was applied.

**Transmission electron microscopy (TEM)**

A high-resolution analytical scanning/ transmission electron microscope (FEI Talos F200X) is operated at 200 keV. It combines high-resolution scanning/transmission electron microscope and TEM imaging with a four-quadrant energy dispersive X-ray spectrometer (EDS) for elemental and compositional mapping. CCPS flakes were exfoliated using scotch tape and transferred into PDMS. After that, the flakes are stamped into the TEM copper grid. Prior to imaging the copper grid was placed in plasma cleaning to remove any contamination left from the stamping process.

**Mode analysis and FDTD simulation**

The electric field profile in the silicon waveguide and the beam propagation were calculated using the MODE Solutions eigenmode solver, a simulator within Lumerical's Device Multiphysics Simulation Suite.

**Spectroscopic imaging ellipsometer**

The optical parameters of multilayer CuCrPS were determined by Accurion's Imaging Ellipsometry (https://accurion.com/company). This system combines optical microscopy and ellipsometry for spatially resolved layer-thickness and refractive index measurements. It is highly sensitive to single- and multi-layer ultrathin films, ranging from mono-atomic or monomolecular layers (sub-nm regime) up to thicknesses of several microns. Additionally, Imaging Ellipsometers perform layer thickness measurements with a spatial resolution down to 1 μm. The ellipsometric parameters (Psi ($\psi$) and Delta ($\Delta$)) were fitted using EP4 model software.

**Stamping of CCPS on the photonic chip**

Multilayered flakes are exfoliated using scotch tape and transferred onto the PDMS substrate. By using the controllable dry transfer method, the selected CCPS flakes are transferred onto the Si waveguides.

**Electrical measurements**

A curve tracer/power device analyzer / (Agilent B1505A) was used to control the biases and measure the I-V characteristics via a pair of standard DC electrical probes under dark conditions.

**Optical characterization**

The optical transmission was tested by edge coupling the light into the device structure through lensed fiber using a tunable laser operating at the SWIR band (Keysight 8164B Lightwave Measurement System). The output response from the devices is collected by an output lensed fiber and detected by a power meter. The light polarization (TE/TM) was calibrated using reference rings fabricated on the same chips with identical geometries. The output optical power intensities were calibrated before the device testing using a standard photodiode power sensor.

**References**

1. Lam, C. F. *et al.* Fiber optic communication technologies: What's needed for datacenter network operations. *IEEE Commun. Mag.* **48**, 32–39 (2010).
2. Han, F. *et al.* Materials with Tunable Optical Properties for Wearable Epidermal Sensing in Health Monitoring. *Adv. Mater.* **34**, 2109055 (2022).

3. Ko, J. H., Yoo, Y. J., Lee, Y., Jeong, H.-H. & Song, Y. M. A review of tunable photonics: Optically active materials and applications from visible to terahertz. *iScience* **25**, 104727 (2022).

4. Ma, Q., Ren, G., Xu, K. & Ou, J. Z. Tunable Optical Properties of 2D Materials and Their Applications. *Adv. Opt. Mater.* **9**, 2001313 (2021).

5. Phare, C. T., Daniel Lee, Y.-H., Cardenas, J. & Lipson, M. Graphene electro-optic modulator with 30 GHz bandwidth. *Nat. Photonics* **9**, 511–514 (2015).

6. Liu, M. *et al.* A graphene-based broadband optical modulator. *Nature* **474**, 64–67 (2011).

7. Datta, I. *et al.* Low-loss composite photonic platform based on 2D semiconductor monolayers. *Nat. Photonics* **14**, 256–262 (2020).

8. Ono, M. *et al.* Ultrafast and energy-efficient all-optical switching with graphene-loaded deep-subwavelength plasmonic waveguides. *Nat. Photonics* **14**, 37–43 (2020).

9. Cheng, Z. *et al.* Phosphorene-assisted silicon photonic modulator with fast response time. *Nanophotonics* **9**, 1973–1979 (2020).

10. Ma, Z., Tahersima, M. H., Khan, S. & Sorger, V. J. Two-Dimensional Material-Based Mode Confinement Engineering in Electro-Optic Modulators. *IEEE J. Sel. Top. Quantum Electron.* **23**, 81–88 (2017).

11. Ross, J. S. *et al.* Electrically tunable excitonic light-emitting diodes based on monolayer WSe2 p–n junctions. *Nat. Nanotechnol.* **9**, 268–272 (2014).

12. Pospischil, A. & Mueller, T. Optoelectronic Devices Based on Atomically Thin Transition Metal Dichalcogenides. *Appl. Sci.* **6**, 78 (2016).

13. Yu, Y. *et al.* Giant Gating Tunability of Optical Refractive Index in Transition Metal Dichalcogenide Monolayers. *Nano Lett.* **17**, 3613–3618 (2017).

14. Chernikov, A. *et al.* Electrical Tuning of Exciton Binding Energies in Monolayer WS 2. *Phys. Rev. Lett.* **115**, 126802 (2015).

15. Ross, J. S. *et al.* Electrical control of neutral and charged excitons in a monolayer semiconductor. *Nat. Commun.* **4**, 1474 (2013).

16. Chen, H. *et al.* Heterogeneous integrated phase modulator based on two-dimensional layered materials. *Photonics Res.* **10**, 1401–1407 (2022).

17. Zhang, Z. *et al.* Racetrack resonator based integrated phase shifters on silicon nitride platform. *Infrared Phys. Technol.* **125**, 104276 (2022).

18. Reed, G. T., Mashanovich, G., Gardes, F. Y. & Thomson, D. J. Silicon optical modulators. *Nat. Photonics* **4**, 518–526 (2010).

19. Noorden, A. F. A., Daud, S. & Ali, J. Implication of plasma dispersion effect for controlling refractive index in microresonator. in 030001 (2017). doi:10.1063/1.4978819.

20. Liu, A. *et al.* High-speed optical modulation based on carrier depletion in a silicon waveguide. *Opt. Express* **15**, 660–668 (2007).

21. Punch, J. Thermal challenges in Photonic Integrated Circuits. in *2012 13th International Thermal, Mechanical and Multi-Physics Simulation and Experiments in Microelectronics and Microsystems* 1/6-6/6 (2012). doi:10.1109/ESimE.2012.6191810.

22. Zhou, J. *et al.* 2D Ferroionics: Conductive Switching Mechanisms and Transition Boundaries in Van der Waals Layered Material CuInP2S6. *Adv. Mater.* **n/a**, 2302419.

23. Lai, Y. *et al.* Two-dimensional ferromagnetism and driven ferroelectricity in van der Waals $CuCrP_2S_6$. *Nanoscale* **11**, 5163–5170 (2019).

24. Wang, C., You, L., Cobden, D. & Wang, J. Towards two-dimensional van der Waals ferroelectrics. *Nat. Mater.* **22**, 542–552 (2023).

25. Ma, R. *et al.* Nanoscale Mapping of Cu-Ion Transport in van der Waals Layered $CuCrP_2S_6$. *Adv. Mater. Interfaces* **9**, 2101769 (2022).

26. Wang, X. *et al.* Electrical and magnetic anisotropies in van der Waals multiferroic CuCrP2S6. *Nat. Commun.* **14**, 840 (2023).

27. Zhu, H. *et al.* Highly Tunable Lateral Homojunction Formed in Two-Dimensional Layered $CuInP_2S_6$ via In-Plane Ionic Migration. *ACS Nano* **17**, 1239–1246 (2023).

28. Park, C. B. *et al.* Observation of Spin-Induced Ferroelectricity in a Layered van der Waals Antiferromagnet $CuCrP_2S_6$. *Adv. Electron. Mater.* **8**, 2101072 (2022).

29. Zhang, D. *et al.* Anisotropic Ion Migration and Electronic Conduction in van der Waals Ferroelectric $CuInP_2S_6$. *Nano Lett.* **21**, 995–1002 (2021).

30. Zhong, Z. *et al.* Robust Threshold-Switching Behavior Assisted by Cu Migration in a Ferroionic $CuInP_2S_6$ Heterostructure. *ACS Nano* **17**, 12563–12572 (2023).

31. Jiang, X. *et al.* Manipulation of current rectification in van der Waals ferroionic CuInP2S6. *Nat. Commun.* **13**, 574 (2022).

32. Rong, D. *et al.* Ion migration induced bipolar diode effect and ferroelectricity modulation in CuInP2S6. *Appl. Phys. Lett.* **122**, 181902 (2023).

33. Deng, J. *et al.* Thickness-Dependent In-Plane Polarization and Structural Phase Transition in van der Waals Ferroelectric $CuInP_2S_6$. *Small* **16**, 1904529 (2020).

34. Wang, Y. *et al.* A gate-tunable artificial synapse based on vertically assembled van der Waals ferroelectric heterojunction. *J. Mater. Sci. Technol.* **128**, 239–244 (2022).

35. Liu, Y. *et al.* Versatile memristor implemented in van der Waals CuInP2S6. *Nano Res.* **16**, 10191–10197 (2023).

36. Li, W. *et al.* A Gate Programmable van der Waals Metal-Ferroelectric-Semiconductor Vertical Heterojunction Memory. *Adv. Mater.* **35**, 2208266 (2023).

37. Ortmann, J. E. *et al.* Ultra-Low-Power Tuning in Hybrid Barium Titanate–Silicon Nitride Electro-optic Devices on Silicon. *ACS Photonics* **6**, 2677–2684 (2019).

38. Mahmoud, M., Cai, L., Bottenfield, C. & Piazza, G. Lithium Niobate Electro-Optic Racetrack Modulator Etched in Y-Cut LNOI Platform. *IEEE Photonics J.* **10**, 1–10 (2018).

39. Cai, L., Mahmoud, A. & Piazza, G. Low-loss waveguides on Y-cut thin film lithium niobate: towards acousto-optic applications. *Opt. Express* **27**, 9794–9802 (2019).

40. Maisonneuve, V. *et al.* Ionic conductivity in ferroic $CuInP_2S_6$ and $CuCrP_2S_6$. *Ferroelectrics* **196**, 257–260 (1997).

41. Selter, S. *et al.* Crystal growth, exfoliation, and magnetic properties of quaternary quasi-two-dimensional CuCrP 2 S 6. *Phys. Rev. Mater.* **7**, 033402 (2023).

42. Maisonneuve, V., Payen, C. & Cajipe, V. B. On CuCrP2S6: Copper Disorder, Stacking Distortions, and Magnetic Ordering. *J. Solid State Chem.* **116**, 208–210 (1995).

43. Susner, M. A., Rao, R., Pelton, A. T., McLeod, M. V. & Maruyama, B. Temperature-dependent Raman scattering and x-ray diffraction study of phase transitions in layered multiferroic CuCr P 2 S 6. *Phys. Rev. Mater.* **4**, 104003 (2020).

44. Asif, M. & Kumar, A. Resistive switching in emerging materials and their characteristics for neuromorphic computing. *Mater. Today Electron.* **1**, 100004 (2022).

45. Yang, Y. *et al.* A new opportunity for the emerging tellurium semiconductor: making resistive switching devices. *Nat. Commun.* **12**, 6081 (2021).

46. Chen, X. *et al.* Optical Transmission Properties of Si3N4 Add-Drop Micro-Ring Resonator Induced by a Fabry–Perot Resonance Effect. *Sensors* **21**, 6370 (2021).

47. Dushaq, G., Paredes, B., Villegas, J. E., Tamalampudi, S. R. & Rasras, M. On-chip integration of 2D Van der Waals germanium phosphide (GeP) for active silicon photonics devices. *Opt. Express* **30**, 15986 (2022).

48. Dushaq, G., Villegas, J. E., Paredes, B., Tamalampudi, S. R. & Rasras, M. S. Anisotropic Van Der Waals 2D GeAs Integrated on Silicon Four-Waveguide Crossing. *J. Light. Technol.* **41**, 1784–1789 (2023).

49. Tamalampudi, S. R., Dushaq, G. H., Villegas, J. E., Paredes, B. & Rasras, M. S. A Multi-Layered GaGeTe Electro-Optic Device Integrated in Silicon Photonics. *J. Light. Technol.* **41**, 2785–2791 (2023).

50. Maiti, R. *et al.* Strain-engineered high-responsivity MoTe2 photodetector for silicon photonic integrated circuits. *Nat. Photonics* **14**, 578–584 (2020).

51. Wei, G., Stanev, T. K., Czaplewski, D. A., Jung, I. W. & Stern, N. P. Silicon-nitride photonic circuits interfaced with monolayer MoS2. *Appl. Phys. Lett.* **107**, 091112 (2015).

52. Gan, S. *et al.* A highly efficient thermo-optic microring modulator assisted by graphene. *Nanoscale* **7**, 20249–20255 (2015).

**Acknowledgement**

The authors are thankful to NYUAD Photonics and Core Technology Platform Facility (CTP) for the analytical, material characterization, devices fabrication, and testing. The first author acknowledges L'Oréal UNESCO For Women in Science Middle East Fellowship.

**Author Contributions**

G.D conceived the research idea, designed the experiments, conducted the majority of the experiments, collected and analyzed the data, and drafted the manuscript, S.S performed the Mode analysis and FDTD simulation, S.T provided insights during the research process, M.R supervised the research project and critically reviewed and revised the manuscript, all authors have read and approved the manuscript.

**Competing Interests**

The authors declare no competing interests.

**Additional information**

Supplementary information

The online version contains supplementary material available at nature website.

**Supplementary Information**

**Unlocking Electro-Optic Tuning in Composite Silicon Photonics Based on Ferroionic 2D Materials**

Ghada Dushaq *, Solomon Serunjogi, Srinivasa R. Tamalampudi, and Mahmoud Rasras*

Department of Electrical and Computer Engineering, New York University Abu Dhabi, P.O. Box 129188, Abu Dhabi, United Arab Emirates

ghd1@nyu.edu, mr5098@nyu.edu

**This file includes:**

**Figure S1.** Structural characteristics: HRTEM of multilayer CCPS

**Figure S2.** Elemental analysis: EDX spectra of multilayer CCPS

**Figure S3.** Electrical characteristics of Au/CCPS/Au structure

**Figure S4.** Optical parameters measurements

**Figure S5.** Mode simulation and confinement factor

**Figure S6**. CCPS integration

**Figure S7.** Optical transmission measurements

Supplement note 1 **optical loss in CCPS using microring resonator**

Supplement note 2 **Change in the effective index of refraction calculation**

**Figure S8.** Experimental setup of electro-optic tuning of CCPS/Si MRR

**Figure S9.** Electro-optic response of composite MRR

**Figure S10.** Resonance shift as a function of applied voltage

**References**

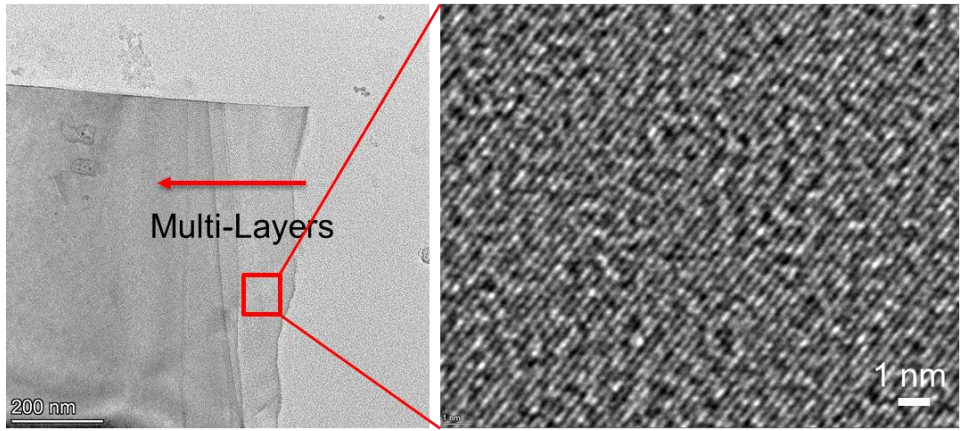

**Figure S1.** High-resolution transmission electron microscopy (HRTEM) of multilayer CCPS captured at the red squared area.

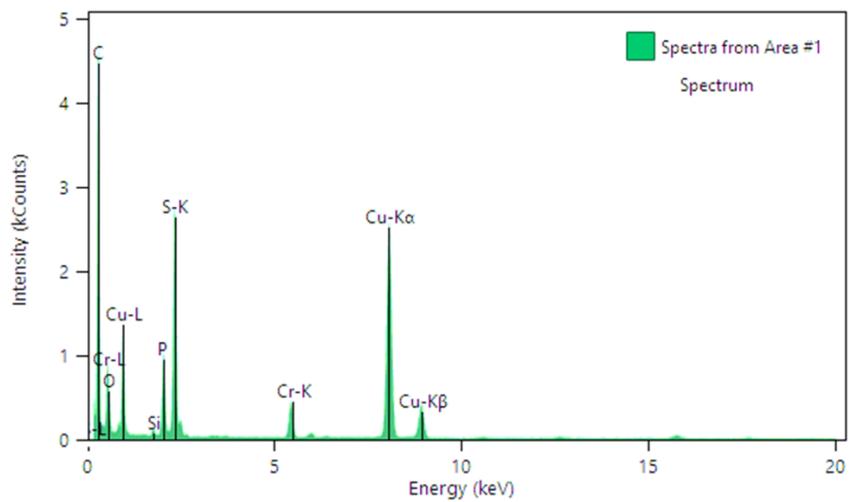

**Figure S2.** Energy dispersive X-ray (EDX) spectra of multilayer CCPS.

**Electrical characteristics of Au/CCPS/Au structure**

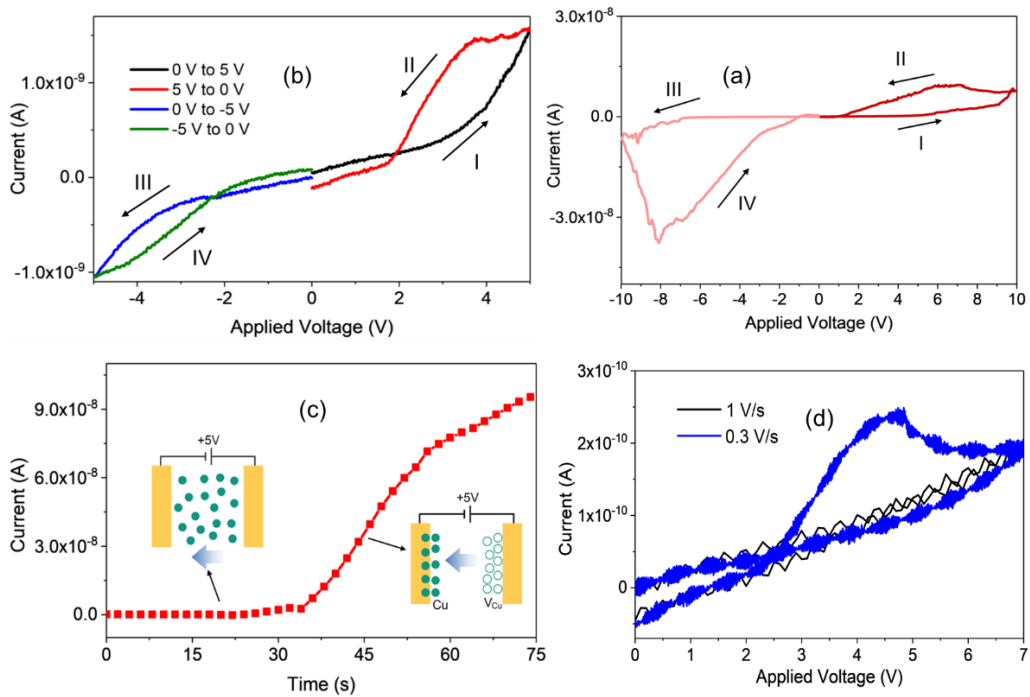

**Figure S3.** Electrical characteristics of Au/CCPS/Au structure (a)-(b) full I−V curves (a) swept in the order 0 V→ 5 V → 0 V → -5 V → 0 V (b) swept in the order 0 V→ 10 V → 0 V → -10V → 0 V labeled as I, II, III, IV (c) current-time (I-t) characteristics at +5 V along with schematic representation of Cu distribution (d) positive I-V cycle characteristics at different voltage ramping rate.

**High-resolution imaging ellipsometry measurements**

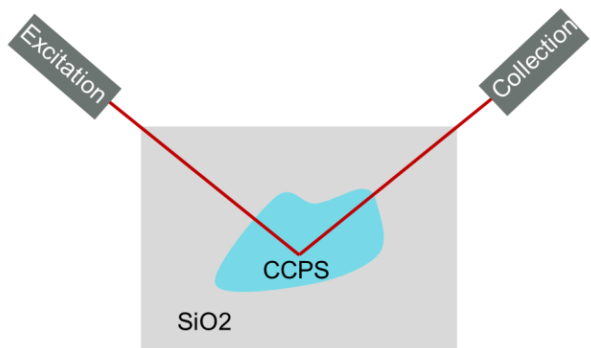

We use spectroscopic ellipsometry on exfoliated crystals of known thickness (measured by AFM) to estimate the optical properties (the refractive index and the extinction coefficient) of CCPS[1–4]. We measure the complex reflectance ratio for CCPS's amplitude ($\Psi$) and phase difference ($\Delta$), which are presented in Figure S4 below. In order to determine the complex refractive index, the experimental data points (black color) are fitted with several analytical models (red color). The bare $SiO_2$/Si substrate located directly next to each investigated layered material was used to collect ellipsometric data in order to reduce substrate-induced uncertainty in the estimation of the complex refractive indices. By using known Si and $SiO_2$ refractive indices to match this data, a measurement of the thickness of the $SiO_2$ layer was obtained. We combine a Lorentzian oscillator-based model with a Tauc-Lorentz oscillator model to explain the exciton resonances. Table S1 provides a description of the fitting parameter.

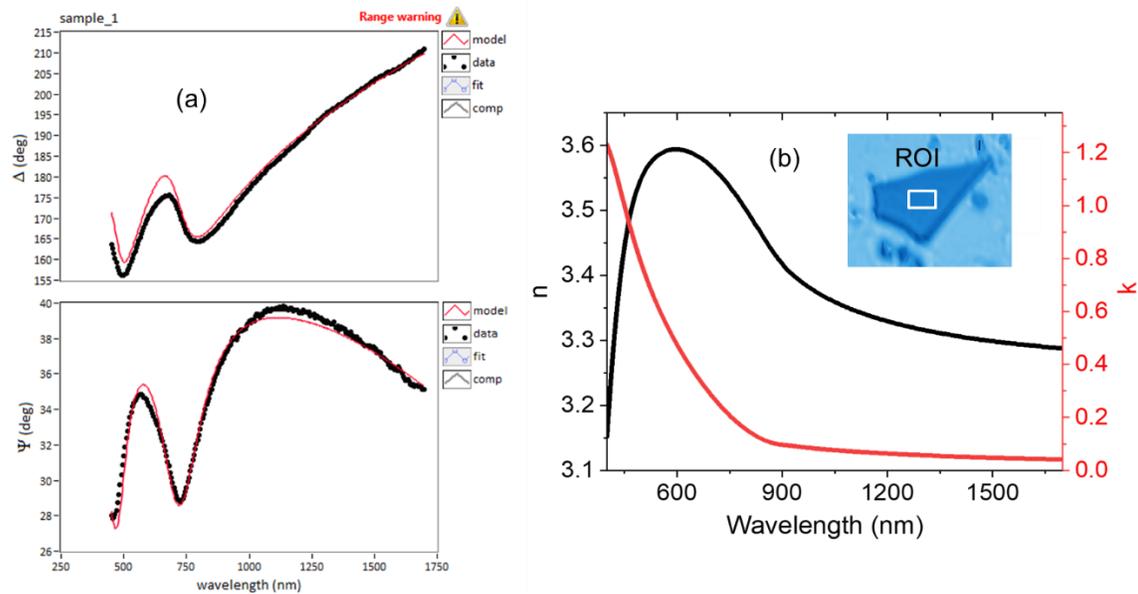

**Figure S4.** Optical parameters extracted from EP4 software (a) ellipsometric spectra of Delta-Psi (b) optical parameters n, k values, inset shows CCPS flake region of interest (ROI).

**Table S1:** Fitting parameters

|  | Best Fit | +/- | Units |
| --- | --- | --- | --- |
| **Thickness** | 62.2 | 0.0 | nm |
| **Eps1** | 5.441 | 0.056 |  |
| **Eps2** | 0.000 | 1.000 |  |
| **Frequency (Lorentz #1)** | 3.214 | 0.013 | eV |
| **Strength (Lorentz #1)** | 53.441 | 1.546 | eV$^2$ |
| **Damping (Lorentz #1)** | 1.641 | 0.028 | eV |
| **Frequency (Lorentz #2)** | 7.936 | 5.995 | eV |
| **Strength (Lorentz #2)** | 10.000 | 1.000 | eV$^2$ |
| **Damping (Lorentz #2)** | 3.000 | 1.000 | eV |
| **A (Tauc-Lorentz)** | 5.000 | 1.000 | eV |
| **E0 (Tauc-Lorentz)** | 5.000 | 1.000 | eV |
| **Gamma** | 1.857 | 2.110 | eV |
| **Eg** | 2.143 | 0.823 | eV |
| **RMSE** | 1.472 |  |  |

**Mode and Beam propagation simulation**

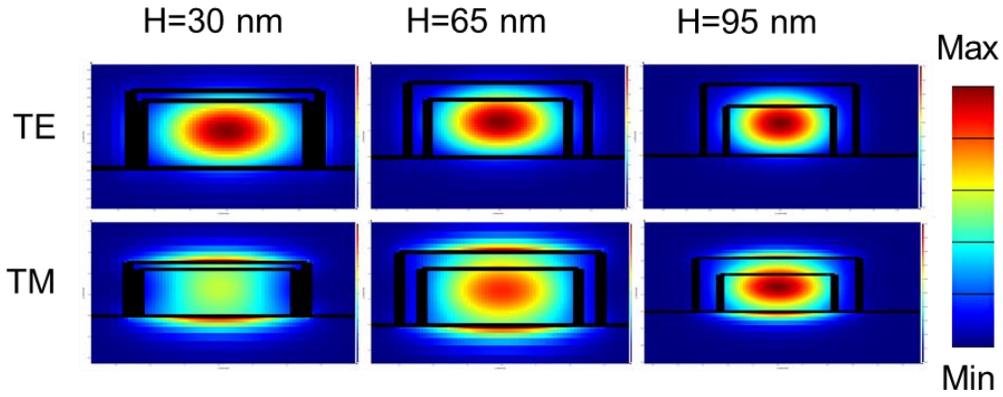

**Figure S5.** Electric-field profiles ($|E|^2$) of TE and TM modes of CCPS with thicknesses of 30 nm, 65 nm, and 95nm on Si at 1310 nm.

**Table S2.** Optical Confinement factor of TE and TM for different CCPS thickness at 1310 nm.

|  | CCPS H=30 nm | | CCPS H=65 nm | | CCPS H=95 nm | |
| --- | --- | --- | --- | --- | --- | --- |
| W=460 nm3 | TE | TM | TE | TM | TE | TM |
| WG+CCPS | 90.7636 | 88.6525 | 92.0503 | 93.4422 | 93.9411 | 95.9502 |
| WG Only | 85.8847 | 85.3944 | 80.6708 | 81.3882 | 71.5104 | 73.3848 |
| CCPS Only | 4.8789 | 3.2581 | 11.3795 | 12.054 | 22.4307 | 22.5654 |

**Bare Waveguide**

| W=460 nm | TE | TM |
| --- | --- | --- |
| WG+CCPS | 0 | 0 |
| WG Only | 87.9009 | 76.7341 |
| CCPS Only | 0 | 0 |

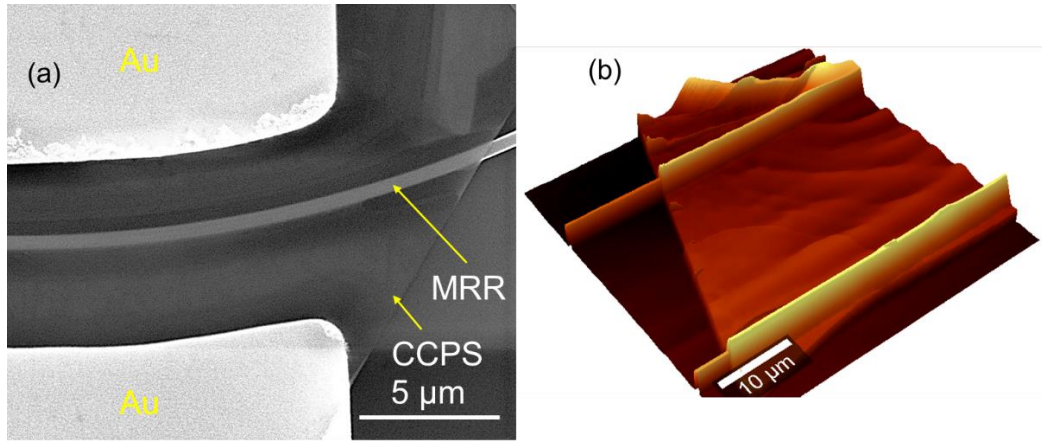

**Figure S6.** CCPS integration (a) SEM image of the transferred CCPS on MRR (b) 3D reconstructed AFM image scan.

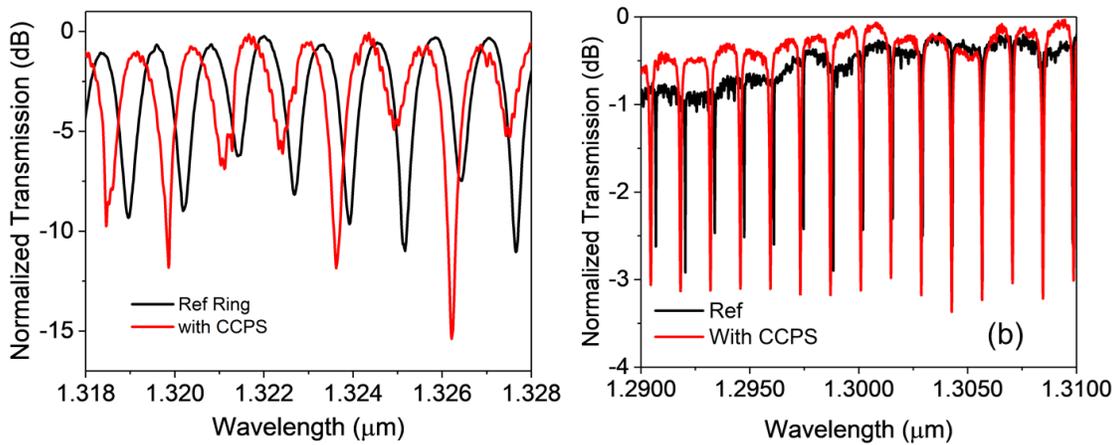

**Figure S7.** Optical transmission measurements (a) transverse magnetic polarization for ~ 65–67 nm and a coverage length of about 30 μm (b) transverse electric polarization for ~ 65–67 nm with 38 μm coverage length.

## 1. Optical losses in CCPS using microring resonator

The change in optical loss per length (Δα) of the ring caused by the multilayer CCPS in the evanescent region is extracted using the reduction observed in the ring's Q-factor. The ring's intrinsic quality factor Q is linked to the loss per unit length as follows[5,6]:

$$Q = \frac{10}{\ln(10)} \frac{2\pi n_g}{\lambda_0 \alpha} \quad (1)$$

Where $\lambda_0$ is the resonance wavelength and $n_g$ is the group index.

As illustrated in Fig. 7b in the main manuscript, the loaded quality factor ($\lambda_0$/FWHM) is extracted from the experimental transmission spectra using Lorentz fitting.

The intrinsic Q in Eq 1 can be calculated from the loaded Q values as follows[7]:

$$Q_i = \frac{2Q_l}{1 \pm \sqrt{T_0}} \quad (2)$$

In which $Q_i$ is intrinsic quality factor, $Q_l$ is loaded Q, and $T_0$ is the normalized transmitted power at the resonance wavelength.

Our chips design is optimized for critical coupling cavity condition at 1310 nm, hence an intrinsic quality factor is ~ $2Q_l$

After CCPS integration, the losses primarily originate from reflections and scattering at the coupling interface between the passive waveguide (air/Si) and the hybrid region, involving mode mismatching losses and flake irregularities. Negligible losses are attributed to material absorption in the SWIR region. The change in the optical loss $\Delta\alpha = \alpha_f - \alpha_i$ can be expressed as[5]:

$$\Delta\alpha = \frac{2\pi n_g}{\lambda_0} \frac{10}{\ln(10)} \left( \frac{1}{Q_f} - \frac{1}{Q_i} \right) \quad (3)$$

**Note**: 10/ln (10) is the conversion from linear to logarithmic.

Here, $Q_f$ is the composite ring quality factor (after integrating CCPS) and $Q_i$ is the bare ring's quality factor (before integration).

The $n_g$ (4.18) value was extracted from the lumerical simulation, the quality factor of the ring before and after was calculated from the measured FWHM of the resonators transmission spectra as represented in Fig.7b using Lorentz fitting. By substituting the measured values obtained for the quality factor before and after integration at a resonant wavelength of 1.30706 µm in equation 3 and using equation 2 to obtain the intrinsic Q, the optical loss due to CCPS can be calculated. This value was then normalized to the CCPS interaction length (2πR/L).

### 2. *Change in the effective index of refraction calculation*

We calculate the change in the real part of the effective index ($\Delta n_{eff}$) of the propagating mode with varying voltage from the change in the resonance wavelength ($\Delta\lambda$) of the transmission response of a ring resonator critically coupled to a bus waveguide (see figure S2) using [8]:

$$\Delta n_{eff} = \frac{\lambda_0 \, \Delta\lambda}{FSR * L}$$

where $\lambda_0$ is the resonance wavelength at 0 V bias, FSR is the free-spectral range (1.4 nm) in terms of wavelength, and L (28 µm) is the CCPS interaction length. **Note:** we calculate the experimental $\Delta n$ by assuming that the phase modulation is entirely due to the Cu migration in CCPS and hence the $\Delta n$ of CCPS equals the $\Delta n_{eff}$ of propagating mode.

**Electro-optic testing setup**

To measure the fabricated device, we use as an input a NIR laser source (Tunable Laser Source Keysight 81606A-113) that is tunable from 1.28 to 1.36 μm and has a 10 kHz linewidth. After sending the beam through a polarization controller, we couple light into the chip through a lensed fiber (side coupling to the chip). The output response from the devices is collected by an output lensed fiber and detected by an optical power meter. A source meter was used to control the biases and measure the I-V characteristics via a pair of standard DC electrical probes under dark conditions.

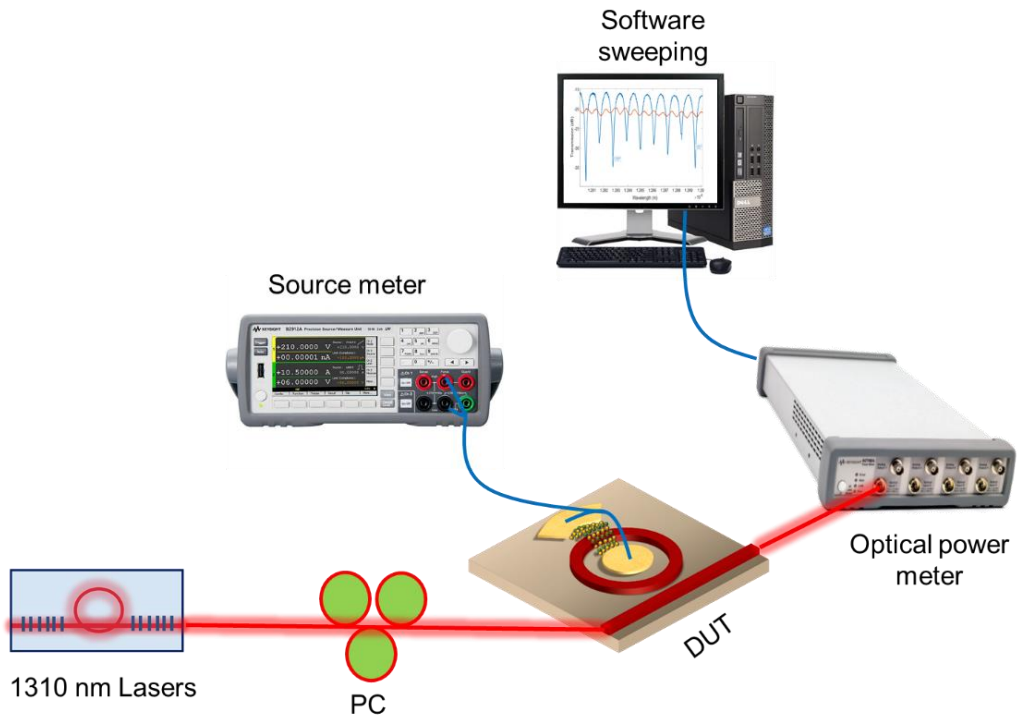

**Figure S8.** Experimental setup of electro-optic tuning of CCPS/Si MRR.

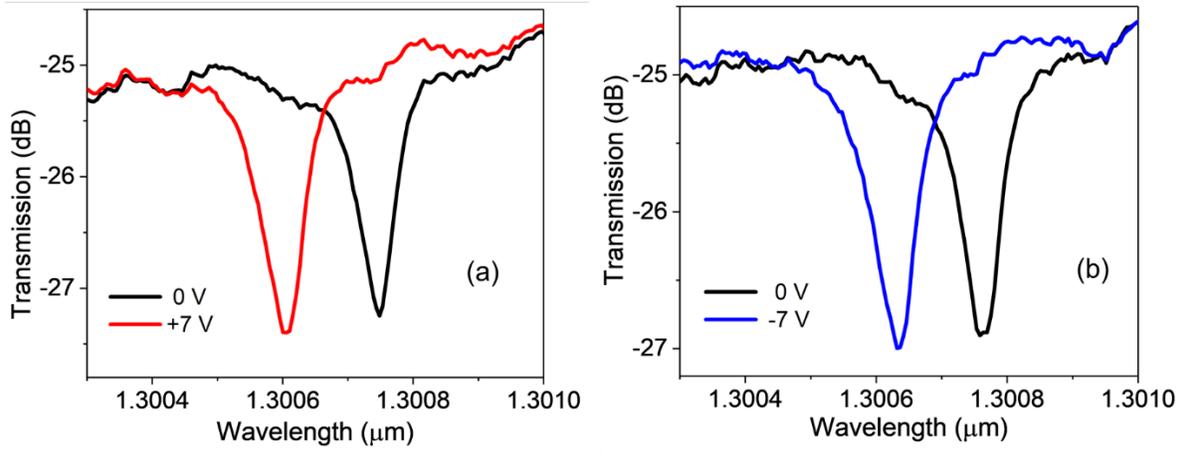

**Figure S9.** The transmission spectra of TE mode of composite MRR at (a) 0V and +7V (b) 0V and -7V

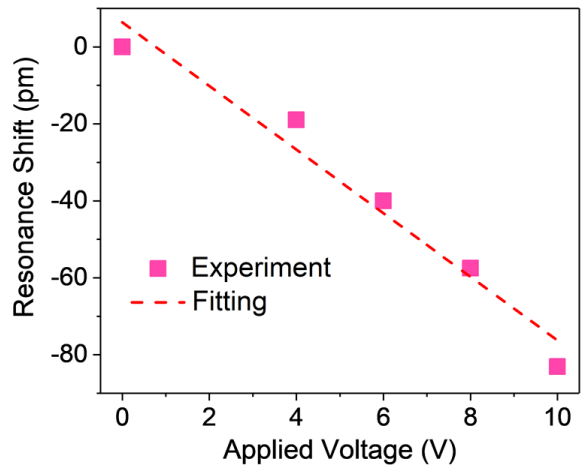

**Figure S10.** The resonance shifts as a function of applied voltage

**References**

1. Zotev, P. G. *et al.* Van der Waals Materials for Applications in Nanophotonics. *Laser Photonics Rev.* **17**, 2200957 (2023).

2. Dushaq, G., Villegas, J. E., Paredes, B., Tamalampudi, S. R. & Rasras, M. S. Anisotropic Van Der Waals 2D GeAs Integrated on Silicon Four-Waveguide Crossing. *J. Light. Technol.* **41**, 1784–1789 (2023).

3. Dushaq, G., Paredes, B., Villegas, J. E., Tamalampudi, S. R. & Rasras, M. On-chip integration of 2D Van der Waals germanium phosphide (GeP) for active silicon photonics devices. *Opt. Express* **30**, 15986 (2022).

4. Tamalampudi, S. R., Dushaq, G., Villegas, J. E., Paredes, B. & Rasras, M. S. A Multi-layered GaGeTe Electro-Optic Device Integrated in Silicon Photonics. *J. Light. Technol.* 1–7 (2023) doi:10.1109/JLT.2023.3237818.

5. Wei, G., Stanev, T. K., Czaplewski, D. A., Jung, I. W. & Stern, N. P. Silicon-nitride photonic circuits interfaced with monolayer MoS2. *Appl. Phys. Lett.* **107**, 091112 (2015).

6. Rabiei, P., Steier, W. H., Cheng Zhang & Dalton, L. R. Polymer micro-ring filters and modulators. *J. Light. Technol.* **20**, 1968–1975 (2002).

7. Miller, S. A. *et al.* Low-loss silicon platform for broadband mid-infrared photonics. *Optica* **4**, 707 (2017).

8. Ortmann, J. E. *et al.* Ultra-Low-Power Tuning in Hybrid Barium Titanate–Silicon Nitride Electro-Optic Devices on Silicon. *ACS Photonics* **6**, 2677–2684 (2019).